\documentclass[aps,longbibliography, twocolumn ,amsmath, amssymb]{revtex4-1}

\usepackage{color}
\usepackage{amsmath}
\usepackage{pifont}
\usepackage{amssymb}  
\usepackage{bbold}
\usepackage{float}
\usepackage{tikz}
\usepackage{makecell}
\usepackage{subfigure}
\usepackage{pifont}   
\usepackage{graphicx} 
\usepackage{dcolumn}  
\usepackage{bm}       
\usepackage{multirow} 
\usepackage{hyperref}
\usepackage{placeins}
\newcommand{\vect}[1]{\mathbf{#1}}

\begin{document}

\renewcommand\floatpagefraction{0.8} 
\renewcommand\topfraction{0.8}       

\author{Michael Vogl}
\thanks{These two authors contributed equally.}
\affiliation{Department of Physics, The University of Texas at Austin, Austin, TX 78712, USA}
\author{Martin Rodriguez-Vega}
\thanks{These two authors contributed equally.}
\affiliation{Department of Physics, The University of Texas at Austin, Austin, TX 78712, USA}
\affiliation{Department of Physics, Northeastern University, Boston, MA 02115, USA}
\author{Gregory A. Fiete}
\affiliation{Department of Physics, Northeastern University, Boston, MA 02115, USA}
\affiliation{Department of Physics, Massachusetts Institute of Technology, Cambridge, MA 02139, USA}
\title{Effective Floquet Hamiltonians for periodically-driven twisted bilayer graphene}
\date{\today}

\begin{abstract}
We derive effective Floquet Hamiltonians for twisted bilayer graphene driven by circularly polarized light in two different regimes beyond the weak-drive, high frequency regime. First, we consider a driving protocol relevant for experiments with frequencies smaller than the bandwidth and weak amplitudes and derive an effective Hamiltonian, which through a symmetry analysis, provides analytical insight into the rich effects of the drive. We find that circularly polarized light at low frequencies can selectively decrease the strength of AA-type interlayer hopping while leaving the AB-type unaffected. Then, we consider the intermediate frequency, and intermediate-strength drive regime. We provide a compact and accurate effective Hamiltonian which we compare with the Van Vleck expansion and demonstrate that it provides a significantly improved representation of the exact quasienergies. Finally, we discuss the effect of the drive on the symmetries, Fermi velocity and the gap of the Floquet flat bands. 
%
\end{abstract}
\maketitle

\section{Introduction}
The recent discovery of strong-correlation effects in twisted bilayer graphene (TBG) generated great interest in moir\'e heterostructures \cite{Wu_2018,Cao2018sc,tsai2019correlated,Codecidoeaaw9770,Yankowitz1059,chichinadze2019nematic,Chou_2019,Guinea13174,PhysRevLett.122.257002,PhysRevB.99.134515,caldern2019correlated,saito2019decoupling,stepanov2019interplay,Kang_2019,Volovik_2018,PhysRevX.8.031089,Ochi_2018,Gonz_lez_2019,Sherkunov_2018,Laksono_2018,Venderbos_2018,PhysRevB.81.165105,salamon2019simulating,PhysRevB.93.035452,Rost_2019,PhysRevB.96.035442,Cheng_2019,Liu_2014,wu2019anomalous,shang2019artificial,Abdullah_2017} and ways to simulate them \cite{salamon2019simulating}. Similar to the behavior in cuprates \cite{Xie_2019,RevModPhys.78.17} at different filling factors superconductivity, Mott-insulating \cite{Codecidoeaaw9770,PhysRevX.8.031089,Cao2018,zhang2020correlated,wong2019cascade} and ferromagnetic behaviour \cite{Sharpe605,Seo_2019} has been observed in TBG. The experimental observations were followed by several theoretical proposals to explain the observations based on the existence of  flat bands which appear at special twist angles~\cite{Bistritzer12233,Wu_2018,PhysRevLett.122.257002}. These flat bands play an essential role for the emergence of strong correlations because the interaction terms become relatively dominant \cite{Kim3364} over the kinetic energy contributions of the dispersive bands~\cite{wong2019cascade,Kim3364,Bistritzer12233,PhysRevB.93.035452,utama2019visualization,saito2019decoupling}. 

In TBG, the flat bands depend strongly on the twist angle between the graphene layers, which is experimentally difficult to set to a precise values. This challenge has lead to several studies proposing different mechanisms to correct for deviations from the magic angle. For example, via pressure \cite{PhysRevB.98.085144, Chittari_2018,Yankowitz2018, Yankowitz1059} or light confined in a waveguide \cite{vogl2020tuning}.

In parallel to the developments on moir\'e lattices, there has been a rapid progress in our understanding of non-equilibrium systems, both experimentally and theoretically, particularly for the case of periodic drives, which may be induced by a laser.  \cite{RevModPhys.83.863,RevModPhys.89.011004,RevModPhys.89.011004,RevModPhys.80.885,RevModPhys.83.1523,Basov2017,doi:10.1146/annurev-matsci-070813-113258,RevModPhys.83.471,doi:10.1080/00018732.2016.1194044,1402-4896-92-3-034004,PhysRevB.81.165433,PhysRevB.93.115420,PhysRevB.95.125401,Iorsh_2017}. The existence of an exponentially-long pre-thermal time regime~\cite{10.1038/nphys4106,PhysRevLett.115.256803,PhysRevB.95.014112,PhysRevLett.116.120401,PhysRevX.7.011026,PhysRevE.93.012130} in driven interacting quantum systems allows one to introduce the notion of effective time-independent theories. The development of several techniques to derive effective Hamiltonians in different drive regimes led to rapid evolution of the Floquet engineering field ~\cite{Eckardt_2015,Blanes2009,Feldm1984,Magnus1954,PhysRevB.95.014112,Bukov_2015,PhysRevA.68.013820,PhysRevX.4.031027,PhysRevLett.115.075301,PhysRevB.93.144307,PhysRevB.94.235419,PhysRevLett.116.125301,PhysRevB.25.6622,PhysRevX.9.021037,Vogl_2019,Vogl2020a,Rodriguez_Vega_2018,Martiskainen2015,rigolin2008,weinberg2015,Jia-Ming2016,verdeny2013,sandovalpaper2019}. For instance, the prediction of an anomalous Hall effect in single-layer graphene driven by circularly polarized light \cite{Oka_2009} has been recently confirmed in experiments~\cite{McIver2020}. More generally, there has been an increased interest in the study of topological transitions induced by periodic drives  \cite{Oka_2009,PhysRevX.3.031005,lindner2011,tong2013,PhysRevB.88.155133,kundu2013,rechtsman2013x,jiang2011,PhysRevLett.107.216601,PhysRevB.89.121401,PhysRevB.90.115423,PhysRevA.91.043625,PhysRevB.91.241404,PhysRevB.97.205415,PhysRevB.97.245401,2019arXiv190902008R,PhysRevB.91.155422,PhysRevB.92.165111,PhysRevB.93.205437,PhysRevLett.116.176401}. 

More recently, the fields of \textit{twistronics} and Floquet engineering crossed paths in twisted bilayer graphene driven by circularly polarized light in free space~\cite{PhysRevResearch.1.023031,li2019floquetengineered,katz2019floquet}. Interesting effects like  topological transitions at large twist angles using high-frequency drives~\cite{PhysRevResearch.1.023031} and the induction of flat bands using near-infrared light in a wide range of twist angles~\cite{katz2019floquet} were found. These studies are mainly numerical, and only provide analytical descriptions in the high drive frequency regime, which we will define rigorously in the next section.

The aim of this work is to derive analytical effective Floquet Hamiltonians that allow us to gain insight into twisted bilayer graphene subjected to circularly polarized light away from the conventional weak drive, high frequency regime of Van Vleck~\cite{2015PhRvA..91c3416P,10.1088/2515-7639/ab387b}, Floquet-Magnus or Brillouin-Wigner approximations~\cite{PhysRevB.94.235419,Mikami_2016}. Our effective Floquet Hamiltonians allow us to elucidate the effects of the interplay of moir\'e lattices and Floquet drives. Particularly, we consider two complementary regimes: i) a regime characterized by weak drive and low frequencies; and ii) a regime characterized by intermediate frequencies and strong drives. The remainder of the manuscript is organized as follows: in Sec. \ref{sec:setup} we describe the system we consider; in Sec.\ref{sec:weak_drive} we examine the low-frequency, weak drive limit; and in Sec.\ref{sec:intermediate_drive} we address the intermediate frequency and intermediate strength drive regime. Finally, in Sec.\ref{sec:conclusion} we present our conclusions and outlook.

\section{System description} 
\label{sec:setup}

\subsection{Static Hamiltonian}
\begin{figure}[t]
	\begin{center}
		\subfigure{\includegraphics[width=8.50cm]{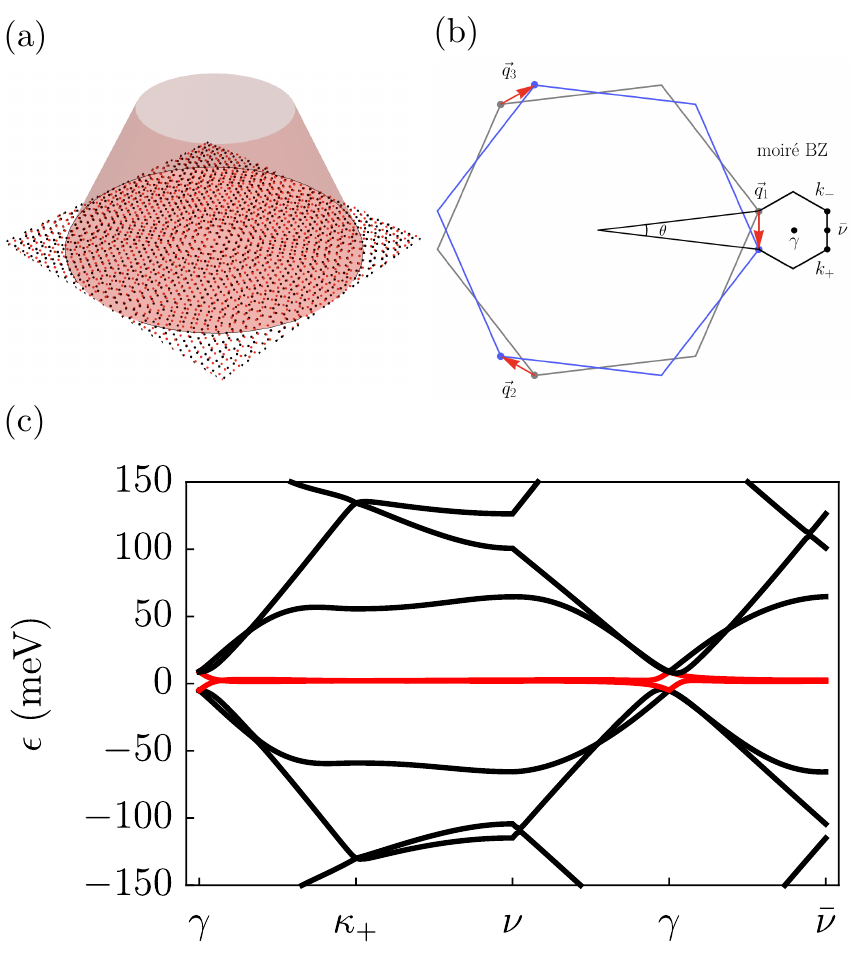}}		
		\caption{(Color online) a) Sketch of twisted bilayer graphene irradiated by circularly polarized light. (b) moir\'e Brilloiun zone. (c) Band structure for twisted bilayer graphene for $w_0=w_1 = 110 $~meV, and $\theta=1.05^\circ$. The low-energy flat bands are highlighted in red.}
		\label{fig:fig1}
	\end{center}
\end{figure}

The starting point of our discussion is the effective Hamiltonian that describes twisted bilayer graphene \cite{Bistritzer12233,Wu_2018,Rost_2019,fleischmann2019perfect,fleischmann2019moir,xie2018nature}
\begin{equation}
\begin{aligned}
&H_{\vect k}(\vect x)=\begin{pmatrix}
h(-\theta/2,\vect k-\kappa_-)&T(\vect x)\\
T^\dag(\vect x)&h(\theta/2,\vect k-\kappa_+)
\end{pmatrix}
\end{aligned},
\label{twist_bilayer_Ham}
\end{equation}
which describes two stacked graphene layers that are rotated with respect to each other by an angle $\theta$, as shown in the sketch of figure \ref{fig:fig1}(a). Here,
\begin{equation}
	h(\theta,\vect k)=\gamma\begin{pmatrix}
	0&f(R(\theta)\vect k)\\
	f^*(R(\theta)\vect k)&0
	\end{pmatrix},
\end{equation}
 is the single-layer graphene Hamiltonian, $f(\vect k)=e^{-\frac{2}{3} i a_0 k_y}+2 e^{\frac{i a_0 k_y}{3}} \sin \left(\frac{a_0 k_x}{\sqrt{3}}-\frac{\pi }{6}\right)$ describes the intralayer hopping amplitude between nearest-neighbor sites, and $\gamma=v_F/a_0$, where we use natural units $\hbar = c = e = 1$. The inclusion of the full structure of $f(\vect k)$ means that this Hamiltonian is valid in the full Brillouin zone and not just near a K point. The interlayer hopping matrix 
 \begin{align}\label{eq:tunn_1}
 	&T(\vect x)=\sum_{i=-1}^1 e^{-i\vect b_i\vect x} T_i,\\
 	&T_i=w_0\mathbb{1}_2+w_1\left(\cos\left(\frac{2\pi n}{3}\right)\sigma_1+\sin\left(\frac{2\pi n}{3}\right)\sigma_2\right),
 	\label{eq:tunn_2}
\end{align}
describes tunneling between the two graphene layers and encodes a hexagonal pattern that has its origin in that the two superimposed graphene lattices which develop a moir\'e pattern (see Fig. \ref{fig:fig1}(b)), where $\vect b_0=(0,0)$, and $\vect b_{\pm 1}= k_\theta\left(\pm \sqrt{3}/2,3/2\right)$ are the reciprocal lattice vectors. Following Refs.~\onlinecite{katz2019floquet,li2019floquetengineered,vogl2020tuning} we introduced an additional parameter $w_1$ into the tunneling term to model relaxation effects, since AB/BA stacking configurations are energetically favoured over AA configurations ~\cite{PhysRevB.96.075311,fleischmann2019moir}. Furthermore, there are indications that $AA$ and $AB$ regions have different interlayer-lattice constants~\cite{PhysRevB.99.205134}. Throughout this work, we fixed $\gamma=v_F/a_0=2.36$ eV, and $a_0 = 2.46\mbox{ \normalfont\AA}$. For a detailed description of the band structure numerical implementation, see the appendix of \cite{vogl2020tuning}. In figure \ref{fig:fig1}(c) we show the band structure for $w_0=w_1 = 110 $~meV, and $\theta=1.05^\circ$, value near the magic angle.

The Hamiltonian in Eq. (\ref{twist_bilayer_Ham}) describes only one valley degree of freedom. A full description of the system would incorporate the two graphene valleys. However, we only consider perturbations induced by light, which cannot induce processes that mix the two valleys. The Hamiltonian in the other valley is connected by a $C_2$ rotation~\cite{Balents2019}. The symmetries of the continuum model Eq. (\ref{twist_bilayer_Ham}) include $C_3$ rotational symmetry about the center of a AA region, $C_2 \mathcal T$ symmetry (taking into account both valleys, the TBG presents time-reversal symmetry $\mathcal T$), and $M_y: y,k_y \to -y,-k_y$ mirror symmetry~\cite{hejazi2019,Balents2019,PhysRevX.8.031089}. In the small-rotation limit, the angle dependence of the graphene sectors can be neglected, leading to an approximate particle-hole symmetry $\mathcal C$~\cite{hejazi2019}.

\subsection{Driven twisted bilayer graphene}

For the driven system, we assume that circularly polarized light is applied in a direction normal to the TBG  plane as sketched in Fig. \ref{fig:fig1}(a). Then, the light enters via minimal substitution as $ k_x \to \tilde k_x(t)  = k_x - A \cos (\Omega t)$, and $k_y \to \tilde k_y = k_y - A \sin (\Omega t)$ leaving the tunneling sector almost unaltered. The reason for this is simple. The inclusion of light in a tight binding model can be done via a Peirls substution for hoppings $t_{ij}\to e^{i\int_{R_i}^{R_j}\vect A d\vect r} t_{ij}$. The interlayer hopping is dominated by hopping between atoms that are almost exactly on top of each other - afterall other atoms are further away and the overlap between orbitals is smaller. Therefore for interlayer couplings mostly longitudinal components of $\vect A$ contribute in the line integral $\int_{R_i}^{R_j}\vect A d\vect r $. Circularly polarized light only has transverse components and therefore has little effect on interlayer couplings. The time-dependent Hamiltonian is
\begin{equation}
    H(\vect x, t) =\begin{pmatrix}
h(-\theta/2,\vect{\tilde k}(t)-\kappa_-)&T(\vect x)\\
T^\dag(\vect x)&h(\theta/2,\vect{\tilde k}(t)-\kappa_+)
\end{pmatrix},
\end{equation}
with $H(\vect x, t+2\pi/\Omega) = H(\vect x, t) $. The Floquet theorem~\cite{Bukov_2015,Eckardt2015a,Mikami_2016} exploits the discrete time-translational symmetry and allows one to write the wavefunctions as $| \psi (t) \rangle = e^{i \epsilon t} |\phi(t) \rangle$, where $|\phi(t + 2\pi/\Omega) \rangle = |\phi(t) \rangle$ and $\epsilon$ is the quasienergy. Replacing $| \psi (t) \rangle$ into the Schr\"odinger equation  leads to $[H(\vect x, t)-i\partial_t] |\phi(t)\rangle = \epsilon |\phi ( t)\rangle $, which governs the dynamics of the periodic system. The exact solution can be generically obtained either by constructing the Floquet evolution operator $ U_F = \text{T} \exp\{-i \int^{2\pi/\Omega}_0  H(s) ds\}=e^{-iH_F T}$ or by employing the extended-state picture. In the extended-state picture, we use the Fourier series $|\phi(t) \rangle  = \sum_n e^{i n \Omega t} |\phi_n \rangle$, which leads to $\sum_m \left(H^{(n-m)} + \delta_{n,m} \Omega m \right) |\phi_m\rangle = \epsilon|\phi_n\rangle$, defined in the infinite-dimensional Floquet-Hilbert space spanned by the direct product of the Hilbert space of the static system and the space spanned by a complete set of periodic functions. The Hamiltonian Fourier modes are given by $H^{(n)} = \int_0^{2 \pi} d\tau/(2\pi) H(\tau) e^{-i \tau n}$, which can be derived by making the replacements 

\begin{widetext}
\begin{equation}
\begin{aligned}
&f(\vect k)\to f^{(n)}(\vect k)= e^{-\frac{1}{3} i (2 a_0 k_y+3 (\theta -\pi ) n)} \left(1+2 e^{i a_0 k_y} \sin \left(\frac{a_0 k_x}{\sqrt{3}}+\frac{2 \pi  n}{3}-\frac{\pi }{6}\right)\right)J_n\left(\frac{2 A a_0}{3}\right)\\
&T(\vect x)\to T^{(n)}(\vect x)=\delta_{n,0} T(\vect x)
\end{aligned},
\end{equation}
\end{widetext}
in Eq. \eqref{twist_bilayer_Ham}.

The two exact approaches outlined above are challenging to use in practice, and one usually has to employ approximations. In the following sections, we will employ a recently developed~\cite{Vogl2020a} approach valid in the weak-drive limit and for arbitrary frequencies. Also, we will introduce improved methods to study the intermediate-amplitude drive regime valid in the high and intermediate frequency regimes.

%
\section{Weak drive regime}
\label{sec:weak_drive}

Thus far, most discussions of twisted bilayer graphene irradiated by circularly polarized light have focused on the high frequency limit. This is for practical reasons because the lower frequency regime, while it is more interesting and relevant for experiments, is also harder to treat using the existing theoretical tools. In Ref.~\onlinecite{Vogl2020a}, we developed a method to address this issue in the weak driving limit. Here, we apply our method using a series of approximations necessary to make progress and gain some analytical insights into the low frequency regime.

If we are interested on the effects of the drive on the low-energy bands, small angles, and weak drives our original Hamiltonian can be approximated with $f(\vect k)\approx f_L( \vect k)= a_0e^{-i\frac{\theta}{2}}(k_x-ik_y)$, in the vicinity of the graphene $K$ point.  The reason we may Taylor expand for small momenta when the twist angle $\theta$ is small is because the moir\'e Brillouin zone is very small i.e. $k_\theta\ll k_D$. Non-linear corrections only become important for higher-energy bands. These higher-energy bands are in turn not relevant for the driven system in the weak-drive limit, since they couple weakly to the low-energy bands.

The time-dependent Hamiltonian within these approximations has the form
\begin{equation}
 H(t)=  H_L+\mathcal Pe^{-i\Omega t}+\mathcal P^\dag e^{i\Omega t},
\label{eq:monocromatic}
\end{equation}
where the monochromatic operator $\mathcal{P} = T^{-1}\int_0^T ds H(\vect x , s) e^{i \Omega s}$ is given by
\begin{equation}
\mathcal P=-A\gamma a_0\begin{pmatrix}
 0&e^{i\theta/2}&0&0\\
 0&0&0&0\\
 0&0&0&e^{-i\theta/2}\\
  0&0&0&0
 \end{pmatrix},
\end{equation}
and $H_L$ is the same as Eq. \eqref{twist_bilayer_Ham} just with $f(\vect k)\to f_L( \vect k)$ linearized momentum dependence.

For weak driving amplitudes $A$ and arbitrary frequency $\Omega$, the periodically driven systems can be described by the effective self-consistent time-independent Hamiltonian~\cite{Vogl2020a} 
 \begin{equation}
H_{\mathrm{eff}}\approx H_L+P\frac{1}{\epsilon-H_L-\Omega }P^\dag+P^\dag\frac{1}{\epsilon-H_L+\Omega} P,
\label{eq:truncfrac}
\end{equation}
where $H_0$ is the time-averaged Hamiltonian, and $\epsilon$ are the quasienergies.  For large frequency drives, we can apply a Van Vleck expansion and obtain the effective Hamiltonian~\cite{2015PhRvA..91c3416P,10.1088/2515-7639/ab387b} $H_{\mathrm{eff}}=H_L+H_\Omega$, where the leading order correction is given by $H_\Omega=-\Delta \tau_0 \otimes \sigma_3$, $\Delta= (A \gamma a_0)^2/\Omega$, and $\sigma_i$, $\tau_i$ are the Pauli matrices in pseudo-spin and layer space, respectively. To keep the notation simple, in the remainder of the text we refer to the approximation $H_{eff}\approx H_L+[P^\dag,P]$ as the Van Vleck approximation.

Therefore, in the high-frequency limit, the main effect is the addition of the gap $\Delta$ in the quasienergy spectrum originating from the breaking of time-reversal symmetry $\mathcal{T}$. This gap is topologically non-trivial, and leads to topological  Floquet  flat  bands with Chern number $C=4$~\cite{katz2019floquet,li2019floquetengineered} which could serve as platforms to realize Floquet fractional Chern insulators~\cite{Grushin2014, li2019floquetengineered}. The relatively large Chern number originates from spin and valley degeneracy~\cite{katz2019floquet,li2019floquetengineered}. 

In order to evaluate the effective Hamiltonian $H_{\text{eff}}$ for arbitrary frequency, we notice that the Brillouin zone has dimensions $k_\theta \propto \sin(\theta/2)$ and therefore the corresponding energy obeys $\hbar v_F k_\theta \ll w_{0,1}$ for sufficiently low angles. This is, for small angles, $T(\vect x)$ introduces the dominant energy scale i.e. $\min\lVert T(\vect x)\rVert\gg \lVert h(\vect k)\rVert$, where $\lVert .\rVert$ is a matrix norm. This estimate can be written more precisely as
\begin{equation}
	\sqrt{(\vect k-\kappa_+)^2+(\vect k-\kappa_-)^2}\ll 3\frac{w_1}{\hbar v_F}
	\label{inequality:small_angle}
\end{equation}
if $w_0\approx w_1$.  (Physically, this is a regime where the interlayer coupling are essential to the physics.)
Therefore, for small enough angles and momenta that fulfill this inequality we may introduce the approximation $\left(\epsilon-H_L\pm\Omega \right)^{-1} \approx \left(\epsilon-H_{T}\pm\Omega \right)^{-1}$ where
\begin{equation}
H_{T}=\begin{pmatrix}
	0&T(\vect x)\\
	T^\dag(\vect x)&0
	\end{pmatrix}.
\end{equation}

Replacing this approximation in the second and third terms of equation (\ref{eq:truncfrac}), we find the effective Hamiltonian $H_{\mathrm{eff}}=H_0+H_\Omega+\mathcal{O}\left(\left(\frac{A}{k_D}\right)^3,\left(\frac{A}{k_D}\right)^2\frac{k_\theta}{k_D}\right)$, where the neglected terms that are third order in small parameters. We find that terms of order $\mathcal{O}\left(\frac{A}{k_D}\right)^4$ vanish. The leading-order correction to the Hamiltonian $H_\Omega$ has the following form
\begin{widetext}
\begin{align}\nonumber
 H_{\Omega}(\vect x)& =  V(\vect x,\Omega) \mathcal \tau_0 \otimes \sigma_0 +  U(\vect x,\Omega)  \tau_3 \otimes \sigma_0  + \frac{1}{2}\Delta_1(\vect x,\Omega)  (\tau_0+\tau_3) \otimes \sigma_3 + \frac{1}{2}\Delta_2(\vect x,\Omega)  (\tau_0-\tau_3) \otimes \sigma_3\\ & +   \delta w_0(\vect x,\Omega) \tau^+ \otimes \sigma_0 +  \delta w^*_0(\vect x,\Omega) \tau^- \otimes \sigma_0 + \beta(\vect x,\Omega)\tau^+ \otimes \sigma_3+ \beta^*(\vect x,\Omega)\tau^- \otimes \sigma_3.
\label{eq:low-freq-corr}
\end{align}
\end{widetext}
%

Equation \eqref{eq:low-freq-corr} is the first main result of our work. The full expressions for each of the terms appearing in Eq. (\ref{eq:low-freq-corr}) are given in appendix \ref{lowfrequHam}, and depend on the quasienergy, which was omitted explicitly for brevity. A perturbative approach generically generates long-range hopping terms as the frequency is arbitrarily decreased. The method here employed leads to the closed form in Eq. (\ref{eq:low-freq-corr}), which contains all the possible terms that can be generated by the drive, even in the low-frequency regime, defined as driving frequency $\Omega\lesssim W$ with $W\sim \mathrm{max}_t\left\lVert H(t)\right\rVert$. Conversely, we define the high-frequency regime for the moir\'e system as $\Omega > W \sim \mathrm{max}_t\left\lVert H(t)\right\rVert$.

Now, we discuss the origin and implications of each the new terms on the symmetries of the system. Due to the assumed approximations, the corrections to the Hamiltonian $H_\Omega (\vect{x})$ presents no momentum dependence and does not commute at different points in space, $\left[H_\Omega (\vect{x}),H_\Omega (\vect{x}') \right] \neq 0$.

\begin{figure}[t]
	\begin{center}
		\subfigure{\includegraphics[width=8.50cm]{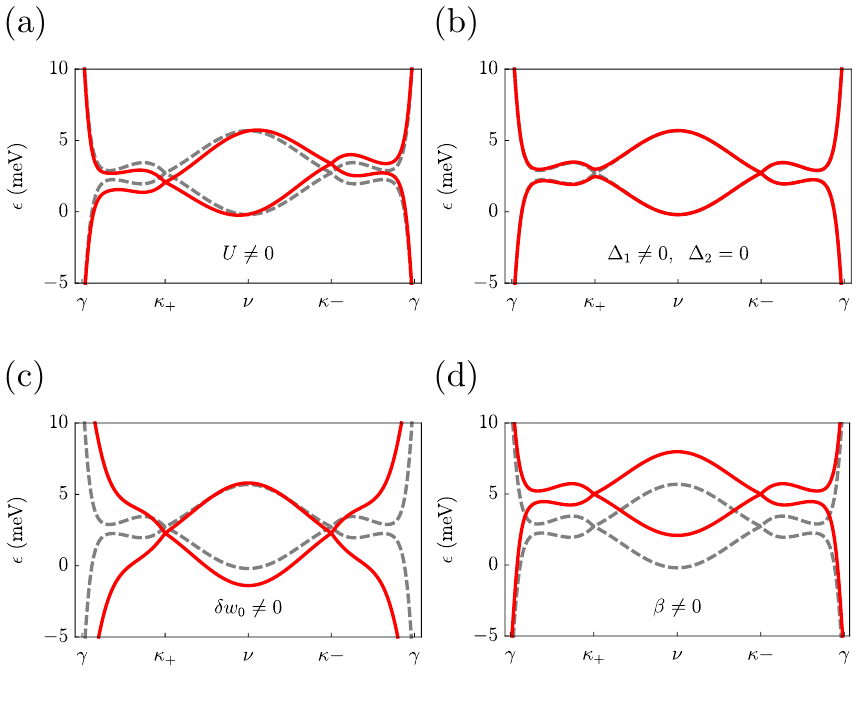}}		
		\caption{(Color online) Sketch of the individual effects of the new term generated by low-frequency and low-intensity circularly polarized light on the  TBG quasienergies. The parameters used are $w_0=w_1 = 110 $~meV, and $\theta=1.2^\circ$. The gray dashed curves correspond to the static case, while the red curve indicates the effect introduced by the non-zero perturbation introduced by light.
		}
		\label{fig:fig2}
	\end{center}
\end{figure}

The first term, $V( \vect x, \Omega ) \mathcal \sigma_0 \otimes \tau_0$, with $V(\vect x, \Omega) \propto \mathcal{O}(\Omega^{-2})$, corresponds to an overall position-dependent potential which does not introduce new physics. The second term, $U(\vect x, \Omega) \sigma_0 \otimes \tau_3$, is a position-dependent interlayer bias with
$ U(x,y)= U(x,-y)$, $ U(x,y)= -U(-x,y)$, and 
$U(x,y) \propto \mathcal{O}(\Omega^{-3})$. This term breaks mirror symmetry $M_y$ and allows a relative shift in quasienergy between the Dirac crossings at $\kappa_{\pm}$, as shown schematically in figure \ref{fig:fig2}(a) for a spatially-uniform constant $U$. Because the $ U(\vect x, \Omega)$ is odd in the $x$-coordinate, $C_2 \mathcal T$ and $C_3$ are also broken when taking the position dependence into account. 

In Bernal-stacked bilayer graphene, a interlayer bias $U$ opens up a gap in the energy spectrum around the $K$ points~\cite{Zhang2009,PhysRevLett.102.256405}. If we introduce a region in space where the sign of the interlayer bias changes, $U \rightarrow -U$, a domain wall forms where the gap inverts, leading to topologically  protected helical (TPH) modes~\cite{Qiao2011,PhysRevLett.100.036804,Alden11256}. In twisted bilayers, even though $U$ does not gap the spectrum, the moir\'e pattern alternating AB/BA regions leads to the formation of topological boundary modes even for spatially-homogeneous interlayer bias $U$~\cite{sanjose2013}. Here, we obtained that circularly polarized light induces an interlayer potential $U(\vect x, \Omega)$ in the low-frequency limit, which could induce the formation of topologically protected helical modes. 

Next, the terms $\Delta_{1/2}(\vect x, \Omega) (\tau_0\pm\tau_3)\otimes \sigma_3 $ with $\Delta_{1/2}(x,-y)=\Delta_{1/2}(x,y)$ and $\Delta_{2}(x,y)=\Delta_{1}(-x,y)$ 
break $M_y$, and $C_2 \mathcal T$ symmetry, which protects the linear band crossing, leading to the opening of a gap at the $\kappa_{\pm}$ points in the mBZ. The $\Delta_{1/2}(\vect x, \Omega)$ position dependence is relevant at order $\mathcal{O}(\Omega^{-3})$, and the asymmetry $\Delta_1 \neq \Delta_2$ is relevant at order $\mathcal{O}(\Omega^{-4})$. When both TBG valleys are taken into account, this term breaks time-reversal symmetry $\mathcal T$  and leads to the formation of topologically non-trivial Floquet flat bands~\cite{li2019floquetengineered,katz2019floquet}. The asymmetry $\Delta_1 \neq \Delta_2$ leads to asymmetric gaps at the $\kappa_{\pm}$ points in the mBZ, as sketched in figure \ref{fig:fig2}(b), where we plot the bands for TBG with a constant term of the form $\Delta_{1}(\tau_0\pm\tau_3)\otimes \sigma_3 $ added. The $\Delta_{1/2}(\vect x, \Omega)$ position-dependence leads to breaking of $C_3$ symmetry. 

The term $\delta w_0(\vect{x}, \Omega) \tau^+ \otimes \sigma_0$ (and its hermitian conjugate) where $\tau^{\pm} = 1/2 \left( \tau_1 \pm i \tau_2 \right)$, $\text{Re } \delta w_0(x,-y) = \text{Re } \delta w_0(x,y)$, $\text{Im } \delta w_0(x,-y) = -\text{Im } \delta w_0(x,y)$, $ \delta w_0(-x,y)=  \delta w_0(x,y)$ introduces a correction to the tunneling amplitude $w_0$, consistent with the symmetries of the static system, except $C_3$. $\delta w_0(\vect{x}, \Omega)$ effectively renormalizes the Fermi velocity at the $\kappa_{\pm}$ points and can modify the position of the magic angles.
To leading order, $\delta w_0(\vect{x}, \Omega)\approx-(A\gamma a_0/\Omega)^2T_{11}(\vect{x}) \cos (\theta )$, where $T_{11}(\vect{x}, \Omega)$ corresponds to the diagonal entry of the tunneling matrix Eq. (\ref{eq:tunn_1}).

In figure \ref{fig:fig2}(d), we schematically show the effect of this term in the Floquet bands. Controlled drive protocols to tune the Fermi velocity of the Floquet zone center flat quasienergy bands have previously been proposed~\cite{vogl2020tuning}. For small angles, large drive frequency $\Omega$ and small quasienergies $\epsilon\ll \Omega$, this term constitutes the second most relevant correction after $\Delta_{1/2}(\vect x, \Omega)$. An accurate description of the quasienergies $\epsilon$ near the Floquet zone center is challenging to achieve with high-frequency expansions such as the Magnus expansion, which highlights the strength of our approach. Crucially, the physics of the Floquet bands near the Floquet zone center is not obfuscated by negligible contributions from static high-energy bands which do not hybridize due to the weak drives considered here. Finally, the correction to the interlayer tunneling $\delta w_0(\vect{x}, \Omega)$ is bears resemblence to effects one would expect from the relaxation of the driven lattice. Particularly, this term only affects the AA-type interlayer coupling $w_0$, which reduces. One could observe a similar effect if the size of AA-type patches were to shrink, which would also lead to the reduction in $w_0$. Secondly if the interlayer distance in AA stacked regions increased, this would also lead to a similar reduction of $w_0$. Therefore the periodic drive is able to mimic these effects.

Finally, we address the term $\beta(\vect{x}, \Omega) \tau^+ \otimes \sigma_3$ (and its hermitian conjugate) with real-space transformation properties  $\beta(-x,y)= \beta(x,y)$, $\text{Re} \beta(x,-y)=\text{Re} \beta(x,y)$, and $\text{Im} \beta(x,-y)=-\text{Im} \beta(x,y)$. To leading order,  $\beta(\vect{x}, \Omega)= i (A \gamma a_0/\Omega)^2 T_{11}(\vect x) \sin \theta + O(\Omega^{-3})$. Neglecting its position dependence, $\beta$ preserves $C_2 \mathcal T$ and $M_y$. Taking the position dependence into account, $\beta(\vect{x})$ breaks both $C_2 \mathcal T$ and $M_y$. Physically $\beta(\vect{x}, \Omega)$ can be interpreted as a pseudo-spin dependent tunneling term.  Therefore, in the weak-drive, small angle and low-frequency regime, circularly polarized light can introduce a collection of symmetry-breaking processes beyond the reach of the high-frequency limit. 

In addition to the small angle limit where Eq.\eqref{inequality:small_angle} is fulfilled let us also consider the opposite limit
\begin{equation}
\sqrt{(\vect k-\kappa_+)^2+(\vect k-\kappa_-)^2}\gg 3\frac{w_1}{\hbar v_F}
\label{inequality:large_angle}
\end{equation}
 where $\left(\epsilon-H_L\pm\Omega \right)^{-1} \approx \left(\epsilon-H_{g}\pm\Omega \right)^{-1}$ with 
\begin{equation}
	H_g=\begin{pmatrix}
h(-\theta/2,\vect k-\kappa_-)&0\\
0&h(\theta/2,\vect k-\kappa_+)
	\end{pmatrix}.
\end{equation}
In this case we find that  $H_{\mathrm{eff}}=H_0+H_\Omega+\mathcal{O}\left(\left(\frac{A}{k_D}\right)^3,\left(\frac{A}{k_D}\right)^2\frac{w_{1,2}}{\gamma}\right)$
\begin{widetext}
\begin{equation}
	H_{\Omega}(\vect k) =  V(\vect k,\Omega) \mathcal \tau_0 \otimes \sigma_0 +  U(\vect k,\Omega)  \tau_3 \otimes \sigma_0  + \frac{1}{2}\Delta_1(\vect k,\Omega)  (\tau_0+\tau_3) \otimes \sigma_3 + \frac{1}{2}\Delta_2(\vect k,\Omega)  (\tau_0-\tau_3) \otimes \sigma_3
\end{equation}
\end{widetext}
      
The gaps are given as
\begin{equation}
	\frac{\Delta_{1/2}(\vect k,\Omega)}{A^2 a_0^2 \gamma ^2}=\frac{ \Omega  \left(\epsilon ^2-\Omega ^2+\left|f_{1/2}(\vect k))\right|^2\right)}{\Lambda_{1/2}(\epsilon,\vect k,\Omega)},
\end{equation}

 the interlayer bias is
\begin{equation}
	\frac{U(\vect k,\Omega)}{A^2 a_0^2 \gamma ^2}=-\frac{\epsilon}{2} \sum_{m=1}^2(-1)^m\frac{    \epsilon^2-\Omega ^2-\left|f_{m}(\vect k))\right|^2}{\Lambda_{m}(\epsilon,\vect k,\Omega)}
\end{equation}

 and 
 \begin{equation}
 	\frac{V(\vect k,\Omega)}{A^2 a_0^2 \gamma ^2}=\frac{\epsilon}{2} \sum_{m=1}^2\frac{    \epsilon^2-\Omega ^2-\left|f_{m}(\vect k))\right|^2}{\Lambda_{m}(\epsilon,\vect k,\Omega)}.
 \end{equation}

where $f_{1/2}(\vect k)=f(R(\mp\theta/2)(\vect k-\kappa_\mp))$, with the property $|f_{1/2}(k_x,-k_y)|^2=|f_{2/1}(\vect k)|^2$,  and
\begin{equation}
	\Lambda_{1/2}(\epsilon,\vect k,\Omega)=\prod_{m=1}^2 \left(\left|f_{1/2}(\vect k))\right|^2-(\epsilon +(-1)^m\Omega )^2\right),
\end{equation}
with $\Lambda_{1/2}(\epsilon, (k_x,- k_y),\Omega) = \Lambda_{2/1}(\epsilon,\vect k,\Omega)$. This property implies that  $V((k_x,- k_y),\Omega) = V(\vect k,\Omega)$, $U((k_x,- k_y),\Omega) = -U(\vect k,\Omega)$, and $\Delta_{1/2}((k_x,- k_y),\Omega) = \Delta_{2/1}(\vect k,\Omega)$. Furthermore,  $U(\vect k,\Omega)$, $V(\vect k,\Omega)$, and $\Delta_{1/2}(\vect k,\Omega)$ are invariant under a $C_3$ rotation of the momentum, since $|f_{1/2}(C_3 \{\vect k \})|^2=|f_{1/2}(\vect k)|^2$. We find that not all terms appearing in Eq.\eqref{eq:low-freq-corr} valid in the limit Eq.\eqref{inequality:small_angle} are generated, and that they are momentum-dependent rather than position-dependent.  A summary of the results for what symmetries get broken by the different terms is given in Table \ref{tab:symmetries}.

\begin{table}[]
\begin{tabular}{|c|c|c|c|}
\hline
                           & \multicolumn{1}{c|}{$C_2 {\cal T}$}      & \multicolumn{1}{c|}{$C_3$} & \multicolumn{1}{c|}{$M_y$} \\ \hline
$U$                        & $\checkmark$                      & $\checkmark$               & $\text{x}$                 \\ \hline
$U(\vect x)$               & $\text{x}$                        & $\text{x}$                 & $\text{x}$                 \\ \hline
$U(\vect k)$               & \multicolumn{1}{c|}{$\checkmark$} & $\checkmark$               & $\checkmark$               \\ \hline
$\Delta$                   & $\text{x}$                        & $\checkmark$               & $\text{x}$                 \\ \hline
$\Delta(\vect x)$          & $\text{x}$                        & $\text{x}$                 & $\text{x}$                 \\ \hline
$\Delta(\vect k)$          & $\text{x}$                        & $\checkmark$               & $\text{x}$                 \\ \hline
$\delta \omega_0$          & $\checkmark$                      & $\text{x}$                 & $\checkmark$               \\ \hline
$\delta \omega_0(\vect x)$ & $\checkmark$                      & $\text{x}$                 & $\checkmark$               \\ \hline
$\beta$                    & $\checkmark$                      & $\text{x}$                 & $\checkmark$               \\ \hline
$\beta (\vect x)$          & $\text{x}$                        & $\text{x}$                 & $\text{x}$                 \\ \hline
\end{tabular}
\caption{This table lists the symmetries that are broken for the different terms that can be generated for the case of position dependence, momentum dependence, or if the term is constant. A checkmark means that the symmetry is preserved, while a cross that symmetry is broken.}
\label{tab:symmetries}
\end{table}

%

The more general case, where neither condition Eq.\eqref{inequality:small_angle} nor the opposite Eq.\eqref{inequality:large_angle} are fulfilled, we can use the general form of $\mathcal{P}$ to find that an effective Hamiltonian has the same structure as Eq.\eqref{eq:low-freq-corr}. However, all terms have an additional momentum dependence (e.g. $\Delta_{1,2}(\vect x,\Omega)\to \Delta_{1,2}(\vect x,\vect k,\Omega)$ etc.). While it is possible to determine that $H_\Omega$ has this structure generally, the coefficients are too cumbersome to compute and are therefore not discussed.

\section{Intermediate drive regime}
\label{sec:intermediate_drive}

\subsection{Issues with the usual form of the rotating frame transformation}

A standard approach for treating systems subjected to intermediately strong drives and intermediate frequencies is applying a rotating frame transformation before the use of a high frequency Magnus expansion \cite{PhysRevX.9.021037,Vogl_2019,Bukov_2015}. To accomplish when a Hamiltonian has the form $H(t)=H_0+\lambda V(t)$, one applies the unitary transformation $U(t)=e^{-i\lambda\int dt V(t)}$ to remove $V(t)$ to lowest order. A large term $\lambda V(t)$  in the Hamiltonian can be traded this way for strongly oscillating terms \cite{Bukov_2015}. This approach allows treating regimes where $\lambda$ is too large for a Magnus approximation to be applicable, and is known to give results that are more reliable than the Magnus expansion \cite{PhysRevX.9.021037,Vogl_2019,Bukov_2015}.

First, we consider the simpler driven Dirac model
\begin{equation}
	H_{\mathrm{D}}=\begin{pmatrix}
	0&k_x-ik_y+\lambda e^{-i\Omega t}\\
	k_x+ik_y+\lambda e^{i\Omega t}&0
	\end{pmatrix},
\end{equation} 
which also describes the upper layer of twisted bilayer graphene near the K point for $w_1=w_0=0$, $\gamma=a_0=1$, $\kappa_{\pm}=0$ and very small $\theta$. 

Application of the unitary transformation $U(t)=e^{-i\lambda\int dt V(t)}$ followed by a zeroth order Magnus approximation leads to a Hamiltonian of the form
\begin{equation}
\begin{aligned}
&H_{\mathrm{eff,D}}=(\vect B+R_y(\tau) (\kappa_1,\kappa_2,0)^T) \cdot \sigma\\
\end{aligned},
\end{equation}
where $R_y(\tau)$ is a rotation matrix around the y-axis by an angle 
\begin{equation}
	\tau=\tan ^{-1}\left(\frac{s J_1\left( s\right)}{s J_0\left(s\right)-  J_1\left(s\right)+\frac{s}{2}}\right),
\end{equation}
$s=\frac{4 \lambda}{\Omega }$, and $J_n(x)$ is the n-th Bessel function of the first kind. The Hamiltonian has a constant field-like part with

\begin{align}
	B_x & =\frac{\lambda}{2}-\lambda \left( J_0\left(s\right)-\frac{ J_1\left(s\right)}{s}\right),\\
	B_y &=0, \\
	B_z & =\frac{1}{s} \lambda \left(J_0\left(s\right)-1\right)-\lambda J_1\left(s\right),
\end{align}

and momenta given by
\begin{align}\nonumber
	\bar k_x& =k_x (s+2 s J_0(s)-2 J_1(s))\\
	&\times\frac{ \sqrt{\frac{2 s^2 J_0(s) (1-2 J_2(s))+s^2 (1-2 J_2(s))+4 \left(s^2+1\right) J_1(s){}^2}{(s+2 s J_0(s)-2 J_1(s)){}^2}}}{2 s},\\
	\bar k_y & =\frac{1}{2} k_y \left(\frac{2 J_1(s)}{s}+1\right).
\end{align}

By inspecting $\bar k_x$ and $\bar k_y$, we realize that $k_x$ and $k_y$ are not treated on equal grounds in this approximation. Specifically, the Fermi velocity has become anisotropic. The quasi-energy spectrum is not rotationally symmetric for large driving $\lambda$. Specifically if we expand $\bar k_{x,y}\approx k_{x,y}\left(1-\frac{ s^2}{16}\mp\frac{ s^4}{384}\right)$ we see that the anisotropic behaviour appears at fourth order in $s$- that is for relatively large $\lambda$. This is in qualitative disagreement with an exact numerical calculations, which present rotationally-symmetric quasi-energies. Since the problem already appears in the Dirac case, we can therefore expect the rotating frame approximation to also produce unphysical artifacts for the more complicated problem of twisted bilayer graphene. It is important to note that the same type of unphysical anisotropy already appears on the level of a first order Magnus expansion \cite{Eckardt_2015}. Therefore, a more careful partial resummation of the Magnus expansion is needed.

\subsection{A better choice of unitary transformation}

In order to avoid introducing unphysical terms in the effective Floquet Hamiltonian, we write the time-dependent Hamiltonian as $H(t)=H_0+\lambda V_1(t)+\lambda V_2(t)$ with $[V_i(t),V_i(t_1)]=0$ and apply the modified unitary transformation $U(t)=e^{-i\int dt V_1(t)}e^{-i\int dt V_2(t)}$ with $V_1(t)=\lambda \cos(\Omega t)\sigma_1$ and $V_2(t)=\lambda \sin(\Omega t)\sigma_2$. There is an associated arbitrariness in the exact form of this unitary transformation arising from the choice of $V_1$ and $V_2$. However, given our implicit Floquet gauge choice $t^*=0$, in a time-ordered exponential that removes all of $V_1+V_2$ we make the smaller error by removing a $V_1$ first, that is the larger of the two at $t^*=0$. In the Dirac model this choice can be justified even better better apostiori by realizing that it restores the rotational invariance in momentum space. 

We will make an analogous choice of unitary transformations for the TBG case in section \ref{sec:Rot_frame_Tw_Bilayer}, where we will explicitly demonstrate that the anisotropy in the Fermi velocity is not present.


\subsection{Improved Van Vleck approximation}

In this section, we identify a procedure to improve the Van Vleck expansion used to obtain an effective Floquet Hamiltonian which we will use as a baseline to compare our improved rotating frame effective Hamiltonian.     


For small twist angles $\theta$ it is sensible to treat $\vect k$ as a small parameter because the dimensions of the moir\'e Brillouin zone are  proportional to $ \sin(\theta/2)$. Therefore, we may approximate $f(\vect k-\vect A)\approx f(-\vect A)+(k_x(\partial_{k_x} f)(-\vect A)+k_y(\partial_{k_y} f)(-\vect A))$. In the weak-strength drive regime, $A \ll a_0$, we employed a simple Taylor expansion. However, in order to capture the effect of stronger drives, we need to improve our approach. For this, we perform a Fourier series in terms of $e^{i\Omega n t}$ instead. The result to first order in Fourier components has the form $f(\vect k-\vect A)\approx a_0(k_x-ik_y)J_0(2a_0A/3)-3J_1(2a_0A/3)e^{i\Omega t}-a_0(k_x+ik_y)J_1(2a_0A/3)e^{-i\Omega t}$. For $2a_0A/3$ not too large compared with unit,  $J_1(2a_0A/3)\ll1$. Therefore, terms like $k_i J_1(2a_0A/3)$ are higher order and can be neglected. We will thus work with the approximation
\begin{equation}
	f(\vect k-\vect A)\approx a_0(k_x-ik_y)J_0\left(\frac{2a_0A}{3}\right)-3J_1\left(\frac{2a_0A}{3}\right)e^{i\Omega t}.
	\label{FOurierseriesApprox}
\end{equation}
This type of approximation is reasonable for small angles and $2a_0A/3\lesssim 1$.

After application of this approximation we can readily improve on the Van Vleck approximation, which we will use to compare our results from the rotating wave approximation. The effective Floquet Hamiltonian keeps the same structure as previously obtained, $H^{\text{vV}}_{\mathrm{eff}}=H_L-\Delta \tau_0 \otimes \sigma_3$ with gap
\begin{equation}
\Delta = \frac{9\gamma^2}{\Omega}J_1\left(\frac{2 A a_0}{3}\right)^2
\label{new_vV_gap}
\end{equation}
and a renormalized Fermi velocity
\begin{equation}
	\tilde v_F=v_F J_0\left(\frac{2 A a_0}{3}\right).
	\label{new_vV_Fermivel}
\end{equation}

\subsection{Rotating frame Hamiltonian}
\label{sec:Rot_frame_Tw_Bilayer}

In this section, we will derive an effective Floquet Hamiltonian using a rotating frame approach, $H^R_{\text{eff}}$, with an improved unitary transformation. Then, we compare the quasienergies obtained with the ones derived from the Van Vleck Hamiltonian $H^{\text{vV}}_{\mathrm{eff}}$.  

We write the time-dependent Hamiltonian for twisted bilayer graphene as $H(t) = H_L + V_1(t) +V_2(t)$, where the time dependent potentials are given as
\begin{align}
	V_1(t)& =-3J_1(2a_0A/3)\cos(\Omega t)\begin{pmatrix}
	\sigma_1^{(-\theta/2)}&0\\
	0&\sigma_1^{(\theta/2)}
	\end{pmatrix}\\
	V_2(t)&=-3J_1(2a_0A/3)\sin(\Omega t)\begin{pmatrix}
	\sigma_2^{(-\theta/2)}&0\\
	0&\sigma_2^{(\theta/2)}
	\end{pmatrix},
\end{align}
where $\sigma_i^{\theta}=e^{-i\frac{\theta}{2}\sigma_3}\sigma_ie^{i\frac{\theta}{2}\sigma_3}$. After applying the unitary transformation $U(t)=e^{-i\int dt V_1(t)}e^{-i\int dt V_2(t)}$ and after taking an average over one period $2\pi/\Omega$ we find the following effective Hamiltonian for twisted bilayer graphene that is subjected to circularly polarized light
\begin{widetext}
\begin{equation}
\hspace*{-0.5cm}
\begin{aligned}
&H^R_{\text{eff}}=R\begin{pmatrix}
(e^{-i\frac{\theta}{2}}\tilde v_F(\vect k-\vect \kappa_-)+\Delta\hat e_z) \cdot \boldsymbol \sigma&\tilde T(\vect r)\\
\tilde T^\dag(\vect r)&(e^{i\frac{\theta}{2}}\tilde v_F(\vect k-\vect \kappa_+)+\Delta\hat e_z)\cdot \boldsymbol \sigma
\end{pmatrix} R^\dagger
\end{aligned},
\label{ROtFramHam}
\end{equation}
\end{widetext}
where $\hat e_z$ is a unit vector in z-direction and $\sigma$ is a vector of Pauli matrices. The unitary transformation
\begin{equation}
R=\left(
\begin{array}{cc}
e^{\frac{3 \gamma  J_1\left(\frac{2 A a_0}{3}\right)}{\Omega} i \sigma_2^{(\theta/2)}} & 0 \\
0 &  e^{\frac{3 \gamma  J_1\left(\frac{2 A a_0}{3}\right)}{\Omega} i   \sigma _y^{(-\theta/2)}} \\
\end{array}
\right),
\end{equation}
allows us to cast the Hamiltonian in more readable form. From this unitary transformation, one can directly identify the origin of the spurious anisotropy in momentum that one would find in a Magnus expansion approach. Particularly, an expansion of $R$ for large frequencies unavoidably leads to such issue.

\begin{figure}[H]
 	\centering
 	\includegraphics[width=1\linewidth]{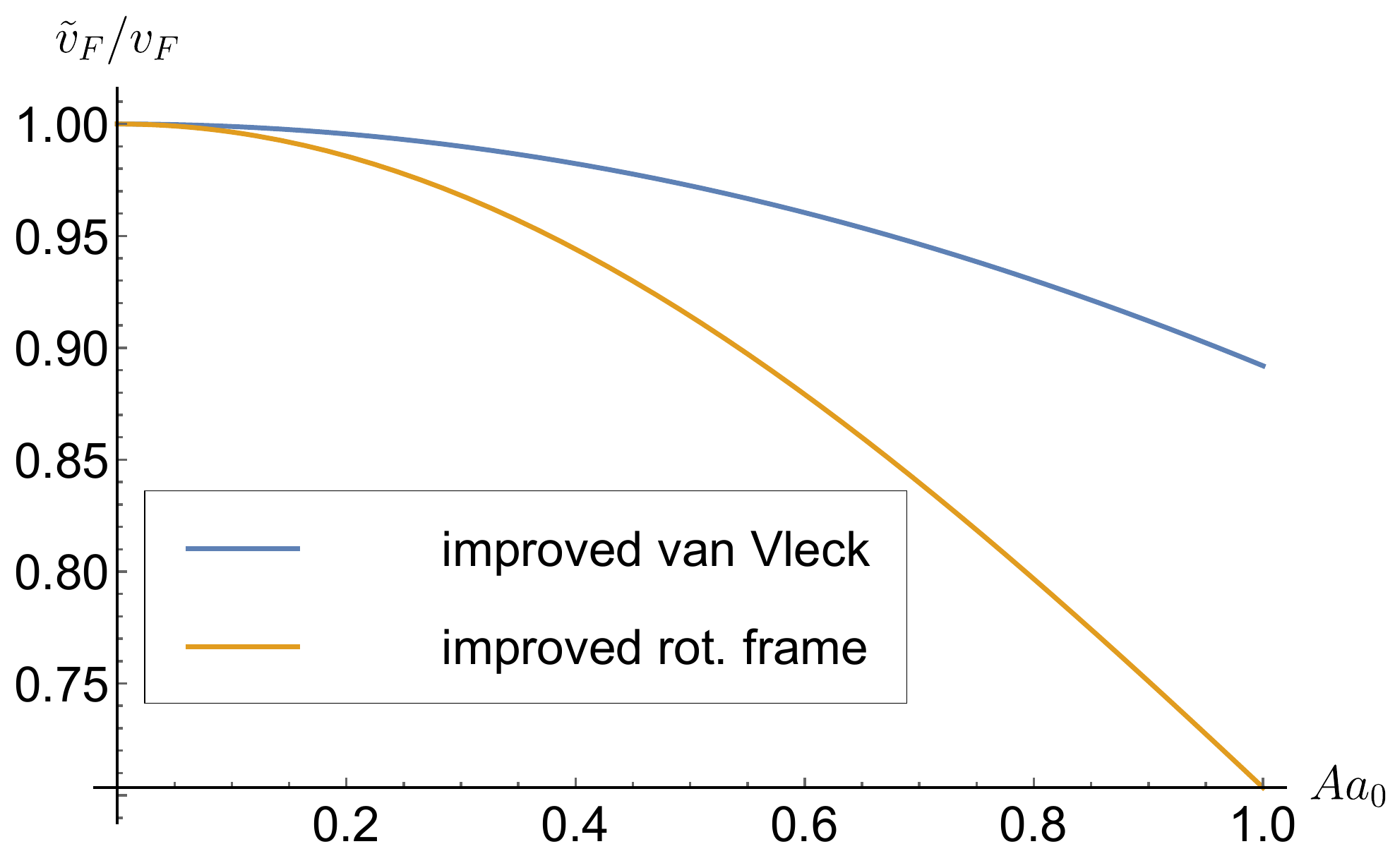}
 	\caption{Fermi velocity normalized to bare $v_F$ for relatively high frequency drives $\frac{\gamma}{\Omega}=\frac{1}{2}$. In blue, we show the Van Vleck result and in orange the renormalized result employing our improved rotating frame approximation. }
 	\label{fig:renormalizedfermivelocity}
 \end{figure}

We find that the Fermi velocity has been renormalized to 
\begin{equation}
	\tilde v_F=v_F J_0\left(\frac{2 A a_0}{3}\right) J_0\left(\frac{6 \gamma  J_1\left(\frac{2 A a_0}{3}\right)}{ \Omega }\right).
\end{equation}
In figure \ref{fig:renormalizedfermivelocity}, we show a plot of the Fermi velocity and compare with to the Fermi velocity from the improved Van Vleck approximation $H^{\text{vV}}_{\text{eff}}$.   We find that the renormalization of the Fermi velocity is $\sim 10\%$ in some regions even for relatively high frequencies.
 
Furthermore, in $H^{R}_{\text{eff}}$, the quasienergy gap that is renormalized to
 \begin{equation}
 	\tilde \Delta=\frac{3}{\sqrt{2}}  \gamma  J_1\left(\frac{2 A a_0}{3}\right) J_1\left(\frac{6 \sqrt{2} \gamma  J_1\left(\frac{2 A a_0}{3}\right)}{\Omega }\right).
 \end{equation}
 
\begin{figure}[H]
 	\centering
 	\includegraphics[width=1\linewidth]{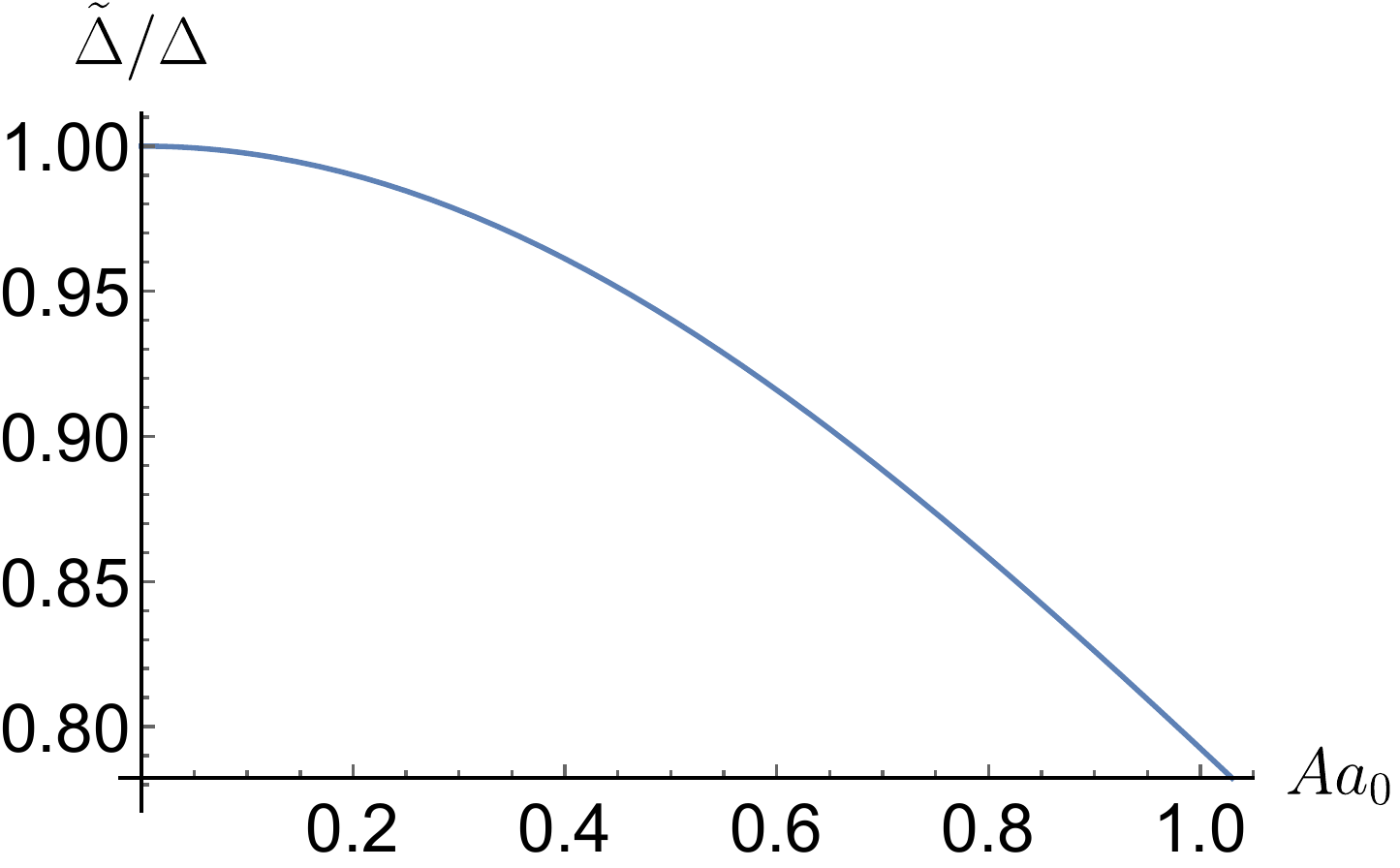}
 	\caption{Ratio $\tilde \Delta/\Delta$ of the renormalized gap $\tilde \Delta$ and the gap of from the Van Vleck $\Delta$ for driving frequency $\Omega=2\gamma$.}
 	\label{fig:renormalizedgap}
 \end{figure}

 A comparison with the Van Vleck result is shown in figure \ref{fig:renormalizedgap}.
 
We find that also in this case that there is considerable difference ($\sim 10\%$ in some regions in parameter space) even for relatively large driving frequencies $\Omega=2\gamma$. 

The most striking difference between $H^{\text{vV}}_{\text{eff}}$ and $H^{\text{R}}_{\text{eff}}$ appears in the tunneling sector, where $H^{\text{R}}_{\text{eff}}$ contains renormalized interlayer hopping
\begin{equation}
 	\begin{aligned}
 	&\tilde T(\vect x)=\sum_{l=-1}^1 e^{-i\vect b_l\vect x} (\tilde T_l-i\beta \sigma_3)\\
 	&\tilde T_n=\tilde w_0\mathbb{1}_2+\tilde w_1\left(\cos\left(\frac{2\pi n}{3}\right)\sigma_1+\sin\left(\frac{2\pi n}{3}\right)\sigma_2\right),
 	\end{aligned}
 	\label{modified_hopping_intermediate_drive}
\end{equation}
with 
\begin{align}
 	\tilde w_1&=w_1 J_0\left(\frac{6 \gamma  J_1\left(\frac{2 A a_0}{3}\right)}{\Omega }\right)\\
 	\tilde w_0&=w_0 \left[1+\sin ^2\left(\frac{\theta }{2}\right) \left(J_0\left(\frac{6 \sqrt{2} \gamma  J_1\left(\frac{2 A a_0}{3}\right)}{\Omega }\right)-1\right)\right],
\end{align}
and a new imaginary term in the AA interlayer coupling
  \begin{equation}
  	\beta=\frac{1}{2} \sin (\theta ) \left(1-J_0\left(\frac{6 \sqrt{2} \gamma  J_1\left(\frac{2 A a_0}{3}\right)}{\omega }\right)\right).
  \end{equation}
In the notation of the previous section \ref{sec:weak_drive} the new coupling term enters as $-i \beta \tau^+\otimes\sigma_3$ and is position dependent. The new dynamically-generated tunneling component $\beta$ breaks $C_3$, the approximate particle-hole symmetry $\mathcal C$, $C_2 \mathcal T$ and reflection $M_y$ symmetries.

\begin{figure*}
	\centering
	\includegraphics[width=1\linewidth]{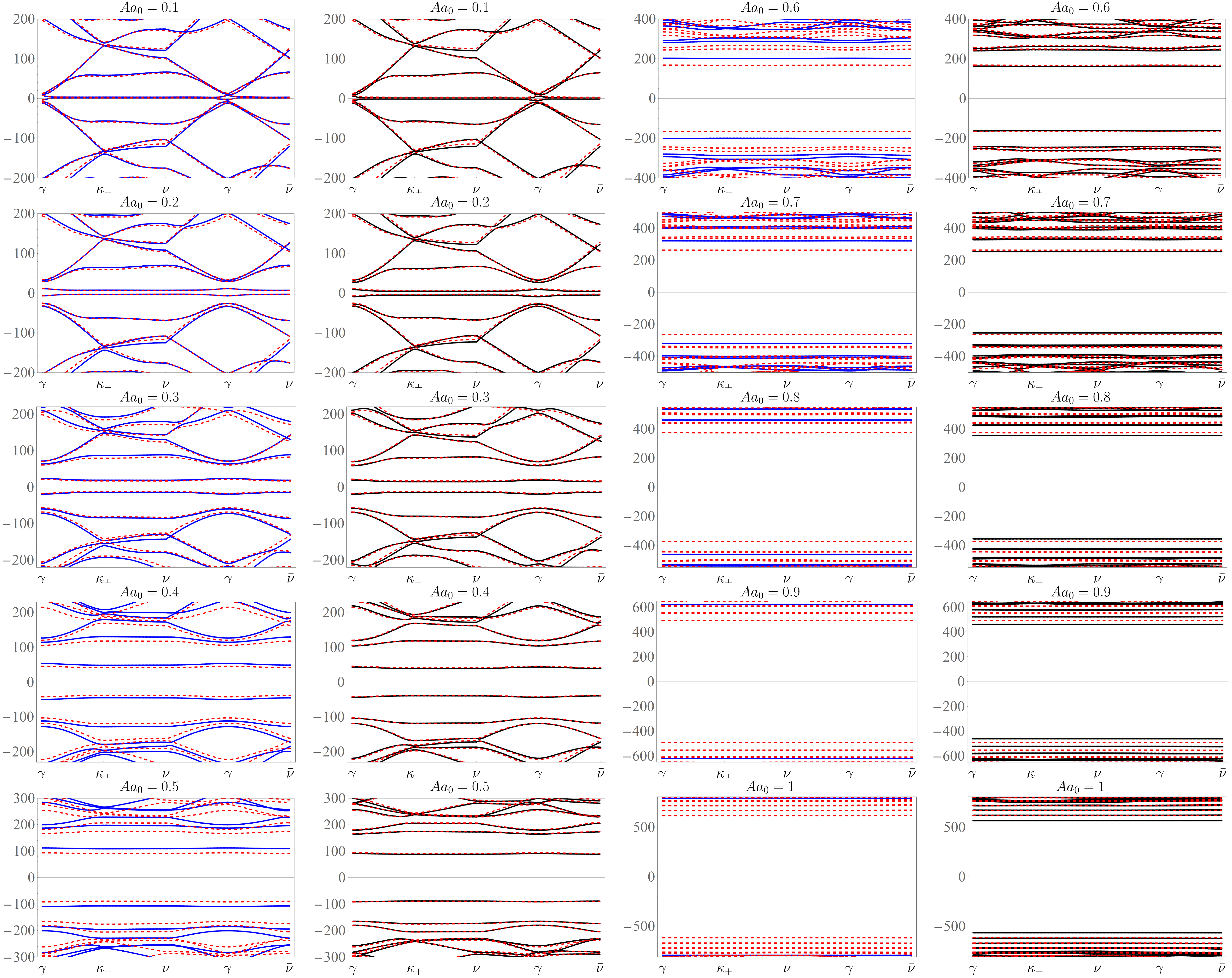}
	\caption{Quasi-energy band structure. The dashed red curves correspond to the exact result, in blue the improved Van Vleck approximation and in black the rotation frame transformation. The parameters used are $\Omega=2\gamma$, $w_1=w_0=110$ meV, $\gamma=2364$ meV and $\theta=1.05^\circ$.}
	\label{fig:bandstructplot}
\end{figure*}

In figure \ref{fig:bandstructplot}, we compare our results using $H^{R}_{\text{eff}}$ in Eq. \eqref{ROtFramHam} to exact numeric results obtained employed an extended space approach \cite{Eckardt_2015}. We use the improved Van Vleck approximation $H^{\text{vV}}_{\mathrm{eff}}=H_L-\Delta \tau_0 \otimes \sigma_3$  as a benchmark.  We find that the Van Vleck approximation is only valid until $a_0A \approx 0.4$, while the new approximation works well until $a_0A\approx 0.8$. The approximation therefore has double the range of validity and therefore is more reliable.

The same observation can be made a bit more lucidly - albeit losing much information- if we compute the relative error of the gap at the $K$ point  $(g_{\mathrm{exact}}-g_{\mathrm{approx}})/g_{\mathrm{exact}}$, where $g_{\mathrm{exact}}$ is the ``exact" numerical gap at the $K$ point and $g_{\mathrm{approx}}$ is the gap for an approximation. For both approximations the result is shown in Fig. \ref{fig:gaprelerrorplot} below.

\begin{figure}[H]
	\centering
	\includegraphics[width=1\linewidth]{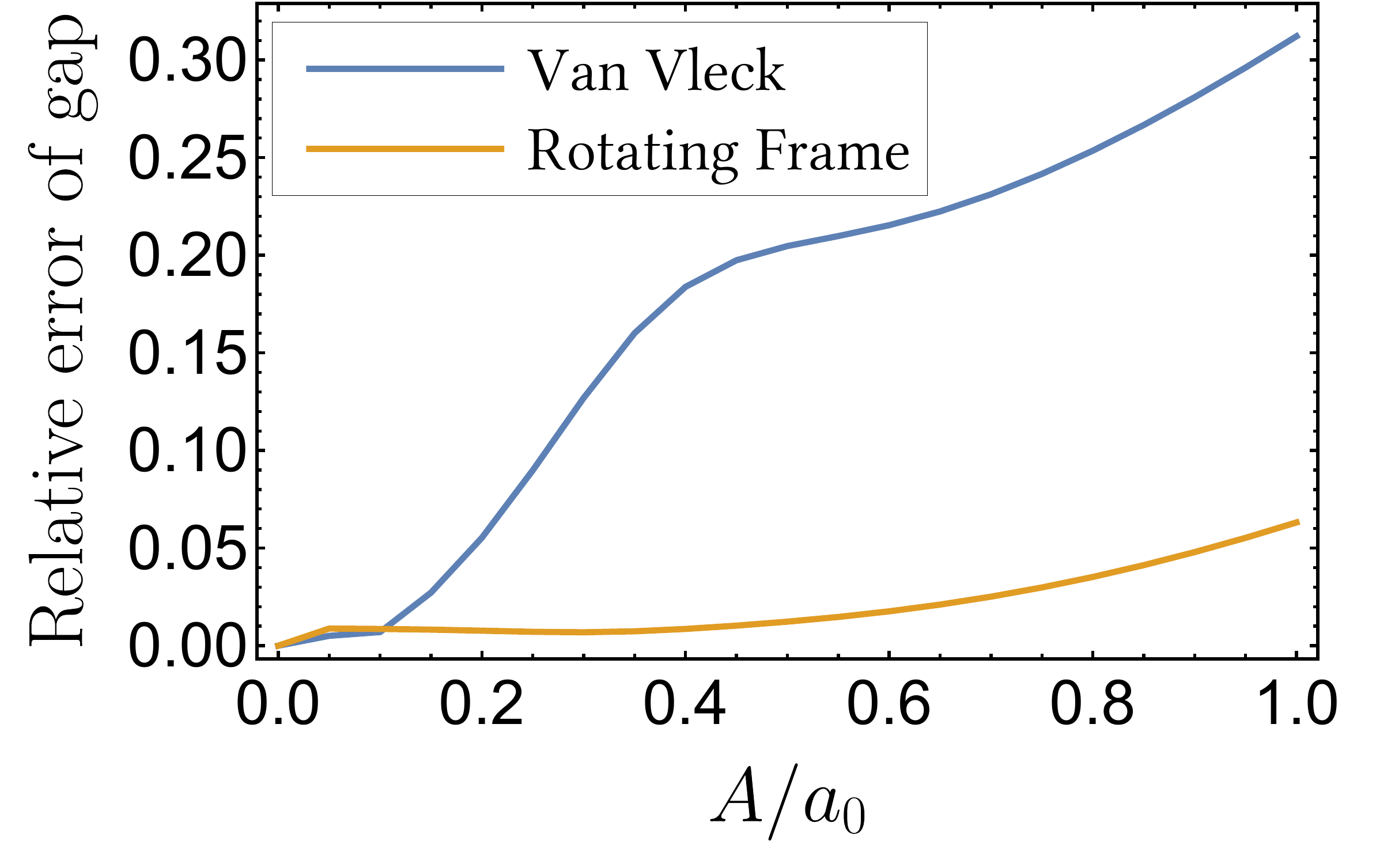}
	\caption{Plot of the relative error for the gap at the $K$ point of the moir\'e Brillouin zone. In blue we present our results from the Van Vleck approximation and in orange the results from our improved rotating frame approximation.}
	\label{fig:gaprelerrorplot}
\end{figure}
It is clear from both plots that the rotating frame approximation  derived in this paper is far more reliable than the Van Vleck approximation.
\section{Conclusion and Outlook}
 \label{sec:conclusion}
 
We have introduced two new effective Floquet Hamiltonians that describe twisted bilayer graphene under the influence of circularly polarized light. The Hamiltonians are applicable in the regimes where the ordinary Van Vleck approximation fails. We found that the weak drive strength Hamiltonian, valid even in the low-frequency regime, gives insight into which new terms a periodic drive can generate well beyond the regime of validity of any other approximation scheme. The usefulness of these scheme is limited by the challenge imposed by the complexity of the terms derived and the self-consistent nature of the low-frequency regime. An important physical effect of the drive in these regime is a renormalization of the interlayer-coupling of the  AA-type. This makes it possible to mimic the effects of some otherwise difficult to achieve structural reorganizations - for instance a change in the distance between the two graphene layers that only appears in AA regions.

The rotating frame Hamiltonian, valid for strong drives and intermediate drive frequencies reveals that the gap at the Floquet zone center, the Fermi velocity, and the interlayer-coupling strengths are renormalized. This effective Hamiltonian is useful for numerical implementation of the quasienergy band structure and posses a wide range of validity. This would make it useful for applications where an extended space calculation may be too expensive. For instance if one studies the effect of disorder additional disorder averages make calculations expensive and therefore it might be more feasible to do these calculations using the effective Hamiltonian we presented rather than resorting to a full treatment in an extended space picture.

\section{Acknowledgements} 

We thank Fengcheng Wu for useful discussions. This research was primarily supported by the National Science Foundation through the Center for Dynamics and Control of Materials: an NSF MRSEC under Cooperative Agreement No. DMR-1720595.  Partial support was from NSF Grant No. DMR-1949701.

 \bibliography{TwistedBilayer}

\appendix
\section{The low frequency Hamiltonian}
\label{lowfrequHam}
 The precise form of the effective low frequency Hamiltonian is given as
 \begin{equation}
 \begin{aligned}
 &	H_{\mathrm{eff}}=H_0+H_\Omega+\mathcal{O}\left(\left(\frac{A}{k_D}\right)^4,\left(\frac{A}{k_D}\right)^2\frac{k_\theta}{k_D}\right)\\&H_\Omega= A^2 \gamma ^2a_0^2\left(
 \begin{array}{cccc}
 W_1^- & 0 & F_- & 0 \\
 0 & W_2^+ & 0 & F_+ \\
 F_-^* & 0 & W_2^- & 0 \\
 0 & F_+^* & 0 & W_1^+ \\
 \end{array},
 \right)\\
 &F_\pm=\frac{e^{\mp i \theta } \left((\epsilon\pm \Omega )^2T_{11}(\vect x)-\text{det}(T(\vect x)) T_{11}^*(\vect x) \right)}{D(\epsilon\pm \Omega)},\\
 &W_n^\pm=-\frac{(\epsilon\pm\Omega )[(\epsilon\pm\Omega)^2-w_0^2 \lambda-w_1^2 \tau_n]}{D(\epsilon\pm \Omega)},\\
 &\lambda=1+4 \cos \left(\frac{\sqrt{3} x_1^\theta}{2}\right) \left(\cos \left(\frac{\sqrt{3} x_1^\theta}{2}\right)+\cos \left(\frac{3 x_2^\theta}{2}\right)\right),\\
 &\tau_n=3-4 \cos \left(\frac{3 x_2^\theta}{2}\right) \sin \left( \frac{\pi}{6} -\frac{\sqrt{3}}{2}  (-1)^n x_1^\theta\right),\\
 &\hspace*{0.8cm}-2 \sin \left((-1)^n\sqrt{3}  x_1^\theta+\frac{\pi }{6}\right);\quad x_i^\theta=x_ik_\theta,\\
 &D(\epsilon)=-\epsilon ^4+\epsilon ^2 \mathrm{Tr}(T^\dag(\vect x)T(\vect x))-\left| \text{detT}\right| ^2,
 \end{aligned}
 \end{equation}
 The quantities in the main text can be derived from here. We find that the intralayer gaps are given as $\Delta_1(\vect x)=\frac{1}{2}(W_1^--W_2^+)$, $\Delta_2(\vect x)=\frac{1}{2}(W_2^--W_1^+)$. A Taylor series reveals
 \begin{equation}
 \begin{aligned}
 &\frac{\Delta_n(\vect x)}{A^2\gamma^2a_0^2}=-\frac{1}{\Omega}-\frac{\epsilon^2+ \mathrm{Tr}(T^\dag T)- w_0^2\lambda  -w_1^2\frac{\tau_1+\tau_2}{2} }{ \Omega^3}\\
 &\hspace*{1.4cm}-(-1)^n\frac{3 \epsilon w_1^2(\tau_1-\tau_2) }{2 \Omega^4}+\mathcal{O}(\Omega^{-5})
 \end{aligned}.
 \end{equation}
 The interlayer bias is given as $U(\vect x)=\frac{1}{4} (W_1^--W_1^+-W_2^-+W_2^+)$. A series expansion is
 \begin{equation}
 	\frac{U(\vect x)}{A^2\gamma^2a_0^2}=\frac{(\tau_1-\tau_2) w_1^2}{2 \Omega^3}+\mathcal{O}(\Omega^{-5}).
 \end{equation}
 As last term from the diagonal block we find the overall potential of form  $V(\vect x)=\frac{1}{4} (W_1^-+W_1^++W_2^-+W_2^+)$, which is expanded as
 \begin{equation}
 \begin{aligned}
 	&\frac{V(\vect x)}{A^2\gamma^2a_0^2}-\mathcal{O}(\Omega^{-5})=-\frac{\epsilon}{\Omega^2}\\
 	&-\frac{\epsilon \left(2 \epsilon^2+6( \mathrm{Tr}(T^\dag T)- \lambda  w_0^2)-3 w_1^2(\tau_1+\tau_2) \right)}{2 \Omega^4}.
 \end{aligned}
 \end{equation}
 Notably the lowest order term is just a constant shift in quasi-energy.
 
On the off-diagonal blocks we find the interlayer hopping strength $\delta w_0(\vect x)=\frac{1}{2}(F_-+F_+)$, which to lower order in $\Omega^{-1}$ is
\begin{equation}
\begin{aligned} 
&\frac{\delta w_0(\vect x)}{A^2\gamma^2a_0^2}-\mathcal{O}(\Omega^{-5})=-\frac{T_{11} \cos (\theta )}{\Omega^2}-\frac{2 i \epsilon T_{11} \sin (\theta )}{\Omega^3}\\
&-\frac{\cos (\theta ) \left(T_{11}(3 \epsilon^2 +\mathrm{tr}(T^\dag T))-T_{11}^*\mathrm{det}(T) \right)}{\Omega^4}
\end{aligned}.
\end{equation}
Furthermore we find that the interlayer hopping has a bias $\beta=\frac{1}{2}(F_--F_+)$, which to low orders has the form
\begin{equation}
	\begin{aligned}
	&\frac{\beta(\vect x)}{A^2\gamma^2a_0^2}-\mathcal{O}(\Omega^{-5})=-\frac{i T_{11} \sin (\theta )}{\Omega ^2}-\frac{2 T_{11} \epsilon  \cos (\theta )}{\Omega ^3}\\
	&-\frac{i \sin (\theta ) \left(T_{11}\left(3 \epsilon ^2+\mathrm{Tr}(T^\dag T)\right)-T_{11}^*\mathrm{det}(T) \right)}{\Omega ^4}
	\end{aligned}.
\end{equation}

\end{document}